\documentclass[11pt]{iopart}
\usepackage{upgreek}

\usepackage[english]{babel}										
\usepackage[protrusion=true,expansion=true]{microtype}				
				
\usepackage[pdftex]{graphicx}									
\usepackage[svgnames]{xcolor}									
\usepackage[hang, small,labelfont=bf,up,textfont=it,up]{caption}	
\usepackage{epstopdf}												
\usepackage{subfig}													
\usepackage{booktabs}												
\usepackage{fix-cm}

\usepackage{txfonts}

\usepackage{amssymb}
\usepackage{txfonts}
\usepackage{color}
\usepackage{hyperref}

\usepackage{soul}

\usepackage{graphicx}

\usepackage{amssymb}

\usepackage{txfonts}
\usepackage{color}
\usepackage{hyperref}
\hypersetup{
  colorlinks = true,
  citecolor = blue,
  urlcolor = blue}

\usepackage[mathscr]{euscript}

\newcommand{{\bfa}}{\mbox{\boldmath$a$\unboldmath}}
\newcommand{{\bfr}}{\mbox{\boldmath$r$\unboldmath}}
\newcommand{{\bfp}}{\mbox{\boldmath$p$\unboldmath}}
\newcommand{{\bfv}}{\mbox{\boldmath$v$\unboldmath}}
\newcommand{{\bff}}{\mbox{\boldmath$f$\unboldmath}}
\newcommand{{\bfF}}{\mbox{\boldmath$F$\unboldmath}}
\newcommand{{\bfA}}{\mbox{\boldmath$A$\unboldmath}}
\newcommand{{\bfchi}}{\mbox{\boldmath$\chi$\unboldmath}}
\newcommand{{\bfmu}}{\mbox{\boldmath$\mu$\unboldmath}}
\newcommand{{\bfnu}}{\mbox{\boldmath$\nu$\unboldmath}}

\newcommand{{\cA}}{\mbox{\boldmath${\cal A}$\unboldmath}}
\newcommand{{\cJ}}{\mbox{\boldmath${\cal J}$\unboldmath}}
\newcommand{{\cF}}{\mbox{\boldmath${\cal F}$\unboldmath}}
\newcommand{{\cG}}{\mbox{\boldmath${\cal G}$\unboldmath}}
\newcommand{{\cE}}{\mbox{\boldmath${\cal E}$\unboldmath}}
\newcommand{{\cB}}{\mbox{\boldmath${\cal B}$\unboldmath}}
\newcommand{{\cX}}{\mbox{\boldmath${\cal X}$\unboldmath}}
\newcommand{{\cY}}{\mbox{\boldmath${\cal Y}$\unboldmath}}
\def\v#1{{\bf#1}}

\newcommand{{\mA}}{\mbox{\boldmath${\mathscr{A}}$\unboldmath}}
\newcommand{{\mZ}}{\mbox{\boldmath${\mathscr{Z}}$\unboldmath}}
\newcommand{{\mF}}{\mbox{\boldmath${\mathscr{F}}$\unboldmath}}
\newcommand{{\mG}}{\mbox{\boldmath${\mathscr{G}}$\unboldmath}}
\newcommand{{\mE}}{\mbox{\boldmath${\mathscr{E}}$\unboldmath}}
\newcommand{{\mB}}{\mbox{\boldmath${\mathscr{B}}$\unboldmath}}
\newcommand{{\mX}}{\mbox{\boldmath${\mathscr{X}}$\unboldmath}}
\newcommand{{\mY}}{\mbox{\boldmath${\cal Y}$\unboldmath}}

\usepackage[top= 2.4 cm, bottom= 2.7 cm, left=2.9 cm, right=2.9 cm]{geometry}

\begin{document}

\title[{\rm J A Heras and R Heras}]{{\fontfamily{qag}\selectfont {\LARGE \textcolor[rgb]{0.00,0.00,0.49}{On Feynman's handwritten notes on electromagnetism and the idea of introducing potentials before fields}}}}

\vskip 30pt
\author{Jos\'e A. Heras$^1$ and Ricardo Heras$^2$}
\address{$^1$Instituto de Geof\'isica, Universidad Nacional Aut\'onoma de M\'exico, Ciudad de M\'exico 04510, M\'exico.
E-mail: herasgomez@gmail.com\\
$^2$Department of Physics and Astronomy, University College London, London WC1E 6BT, UK. E-mail: ricardo.heras.13@ucl.ac.uk }
\begin{abstract}
\noindent In his recently discovered handwritten notes on ``An alternate way to handle electrodynamics'' dated on 1963, Richard P. Feynman speculated with the idea of getting the inhomogeneous Maxwell's equations for the electric and magnetic fields from the wave equation for the vector potential. With the aim of implementing
this pedagogically interesting idea, we develop in this paper the approach of introducing the scalar and vector potentials before the electric and magnetic fields. We consider the charge conservation expressed through the continuity equation as a basic axiom and make a heuristic handle of this equation to obtain the retarded scalar and vector potentials, whose wave equations yield the homogeneous and inhomogeneous Maxwell's equations. We also show how this axiomatic-heuristic procedure to obtain Maxwell's equations can be formulated covariantly in the Minkowski spacetime.
\vskip 5pt
 ``\emph{He (Feynman) said that he would start with the vector and scalar potentials, then everything would be much simpler and more transparent.}''

 \rightline {M. A. Gottlieb-M. Sands Conversation\footnote[1]{M. A. Gottlieb comments that ``In 2008 Matt Sands told me that in about the middle of the 2nd year of the FLP lectures [Feynman Lectures on Physics], Feynman started to complain that he was disappointed that he had been unable to be more original. He explained that he thought he had now found the `right way to do it' -- unfortunately too late. He said that he would start with the vector and scalar potentials, then everything would be much simpler and more transparent. These notes [the five handwritten pages dated on 1963] are the only known documentation of Feynman's `right way to do it.''' Extract taken from Feynman Lecture Notes of M. A. Gottlieb appearing in the webpage: \href{http://www.feynmanlectures.caltech.edu/info/notes.html}{http://www.feynmanlectures.caltech.edu/info/notes.html}. To put  these comments in context, we must say that Feynman was not satisfied with the standard presentation of electromagnetism appearing in the second volume of Feynman's Lectures \cite{1}. Regarding this presentation he wrote: ``I couldn't think of any really unique or different way of doing it --or any way that would be particularly more exciting than the usual way of presenting it. So I don't think I did very much in the lectures on electricity and magnetism.''}}
\end{abstract}
\vskip 30pt

\section*{{{\fontfamily{qag}\selectfont {\large \textcolor[rgb]{0.00,0.00,0.49}{1. Introduction}}}}}
Searching through the historical Caltech archives,  Gottlieb \cite{2} recently discovered five handwritten pages of notes dated on  Dec. 13, 1963 in which Richard P. Feynman sketched some ideas on an alternate way to handle electrodynamics. More recently, De Luca et al. \cite{3} have presented their version of how a part of Feynman's ideas may be implemented so that they may be used as a supplementary material to usual treatments on electrodynamics. Following Feynman's ideas to a certain extent, they  heuristically obtained the Lorentz force and the homogeneous Maxwell's equations. Their procedure can be briefly outlined as follows.
\begin{itemize}
  \item Following Feynman, De Luca et al. \cite{3} assume that the force on an electric charge $q$ moving with velocity
$v_j$ is of the generic form $F_i=q(E_i+v_jB_{ij})$, where $E_i$ and $B_{ij}$ are functions of space and time to be determined (summation on repeated indices is understood). Next, this 3-force is assumed to be the spatial component of a 4-force. Considering the relativistic transformation of this 4-force, the form of the 3-force is found to be: $\mathbf{F}=q(\mathbf{E}+\mathbf{v}\times \mathbf{B})$ where $\textbf{B}$ represents the independent components of $B_{ij}$ (they make $c\!=\!1$).  Through this procedure the relativistic transformations of the vectors $\mathbf{E}$ and $\mathbf{B}$  may be identified with those of the electric and magnetic fields and this leads to the conclusion that $\mathbf{F}=q( \mathbf{E}+\mathbf{v}\times \mathbf{B})$ is the Lorentz force.  We should emphasize  that this procedure to obtain the Lorentz force was
roughly sketched out by Feynman in his handwritten notes. In our opinion, however, Feynman's route  to the Lorentz force is criticisable:
The hypothesis of a force linear in the velocity is not sufficiently well justified. But we must also recognize that the derivation of a Lorentz-like force from relativistic considerations and the assumption of a force depending linearly on velocity are conceptually interesting.

  \item De Luca et al. \cite{3} assume the relativistic action $S=\int_{t_1}^{t_2}[-m_0ds-qA_\mu dx^\mu]$, where $A_\mu$ is the 4-potential (they now use relativistic notation). They then
   vary this action to find the force $\mathbf{F}= q[-\nabla\Phi-\partial\mathbf{ A}/\partial t+\mathbf{v}\times (\nabla \times \mathbf{A})]$.
   Comparison of this force with the previously obtained Lorentz force yield the relations $\mathbf{E}=-\nabla\Phi-\partial \mathbf{A}/\partial t$ and $\mathbf{B}=\nabla\times \mathbf{A}$ which imply the homogeneous Maxwell's equations $\nabla\cdot \textbf{B}=0$ and $\nabla \times \mathbf{E}=-
   \,\partial \mathbf{B}/\partial t$. This procedure based on the least action principle, which starts with potentials and ends with the homogeneous Maxwell equations, was not drawn in Feynman's  handwritten notes. In getting the homogeneous Maxwell equations,
   De Luca et al. \cite{3} considered the Feynman's Hughes Lectures \cite{4}. They justify their procedure by arguing that ``It is conceivable that Feynman had something like this in mind in 1963, when he wrote his notes.''\footnote[2]{This opinion of De Luca et al. is questionable. In a 1966 interview with C. Weiner [see the website: \href{https://www.aip.org/history-programs/niels-bohr-library/oral-histories/5020-1}
   {https://www.aip.org/history-programs/niels-bohr-library/oral-histories/5020-1}] Feynman said: ``I've now cooked up a much better way of presenting
the electrodynamics, a much more original and much more powerful way than is in the
book.'' Nevertheless, the Lagrangian approach leading to the Lorentz force used by the Luca et al. \cite{3} was well-known in the 1960's. It is hard to believe that for those years Feynman would refer to this Lagrangian approach by claiming that it was ``much more original.'' Furthermore,
according to Dyson (see Ref.~\cite{37}) this Lagrangian approach was well-known by Feynman at around 1948. Therefore, it is conceivable that Feynman in 1963 had in mind an original explanation different from the Lagrangian explanation when he wrote his notes.}
   \end{itemize}
Although the attempt of De Luca et al. \cite{3} to make useful Feynman's alternate way to handle electrodynamics is valuable, it turns out to be incomplete because the inhomogeneous Maxwell equations: $\nabla\cdot \textbf{E}=\rho/\epsilon_0$ and $\nabla \times \mathbf{B}=\mu_0\textbf{ J} +\epsilon_0\mu_0\partial \mathbf{E}/\partial t$ were not inferred. De Luca et al. recognize this incompleteness but they make no attempt to address this problem. Interestingly, Feynman himself wasn't sure how to get the inhomogeneous Maxwell's equations as may be seen in the third point of his first handwritten page, which is partially reproduced in figure~1. With signs of doubt (he wrote: How!?) he speculated with the idea that such inhomogeneous equations could be obtained from the wave equation for the vector potential or from ``other principle.'' It is not surprising that Feynman was interested in following the unconventional route of starting with potentials before considering the inhomogeneous Maxwell's equations. Feynman liked the idea that potentials and fields had the same level of reality. In the context of the Aharonov-Bohm effect and referring to the vector potential $\mathbf{A}$ and the magnetic field $\mathbf{B}$, he wrote \cite{5}: ``$\mathbf{A}$ is as real as $\mathbf{B}$-realer, whatever that means.''

We think that the speculative idea raised by Feynman of introducing potentials before fields is pedagogically interesting and deserves to be explored. In this sense it is pertinent to say that in the traditional presentation of Maxwell's equations appearing in textbooks, potentials are introduced using the homogeneous Maxwell's equations. The electric and magnetic fields expressed in terms of the scalar and vector potentials are then used in the inhomogeneous Maxwell's equations, obtaining explicit retarded forms of these potentials. The reversed idea of introducing first retarded potentials satisfying wave equations and then deriving the homogeneous Maxwell's equations does not seem to have been explored so far, at least in the standard literature available to us. However, we believe that the idea exploring alternative presentations of Maxwell's equations is important for pedagogical and conceptual reasons.

\begin{figure}
  \centering
  \includegraphics[width= 350pt]{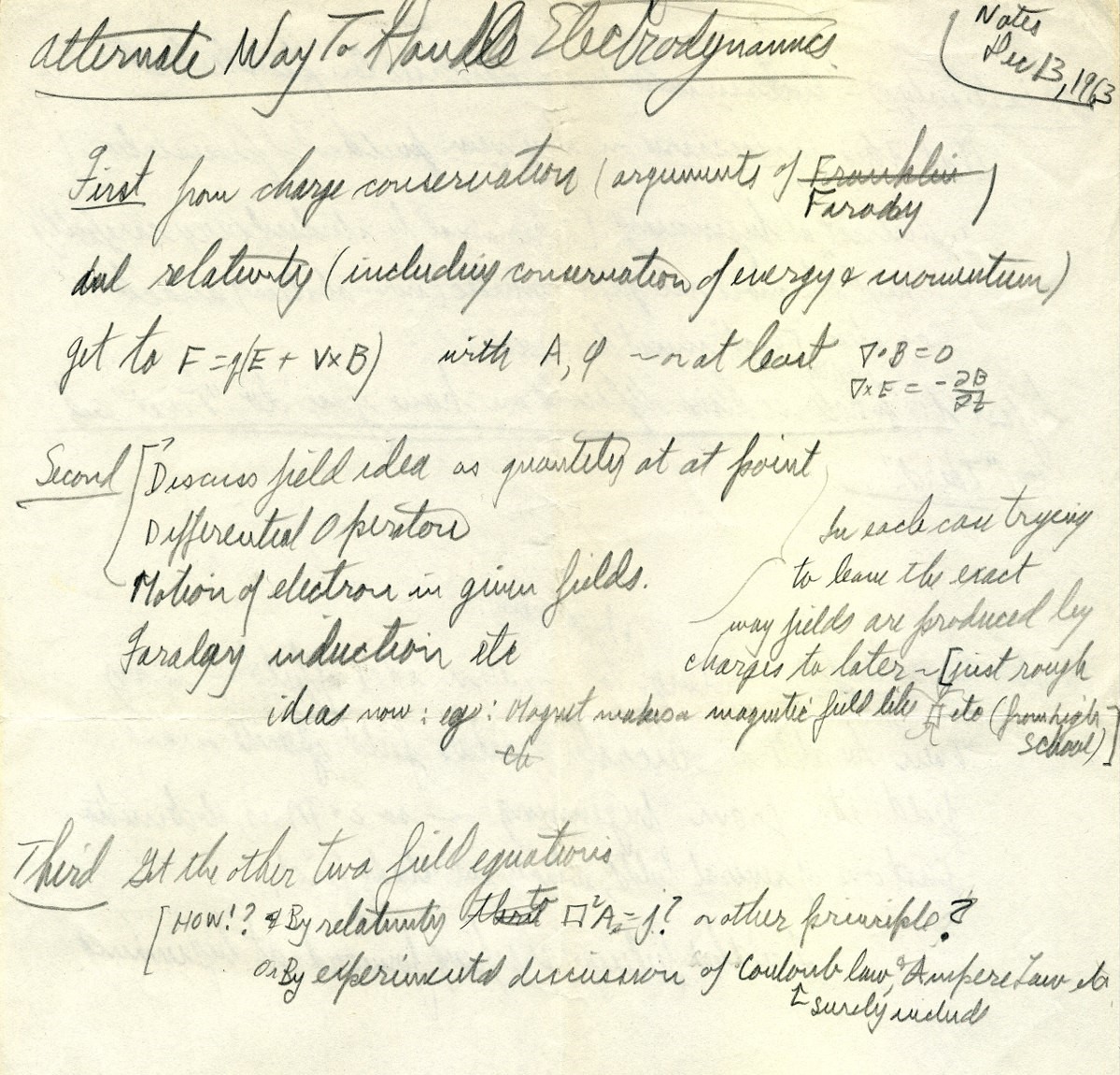}
  \caption{Extract of Feynman's first handwritten page entitled ``Alternate	Way	to	Handle	Electrodynamics.'' The point 3 states:
  ``\underline{Third}  Get the other two field equations .... HOW!? By relativity \textst{thru} (to) $\Box^2 A=j$? or other principle? ....
  Or By experimental discussion	of Coulomb law (surely include), Ampere Law etc.'' Reproduced with the permission of the estate of Richard P. Feynman and The California Institute of Technology.}
\label{Fig1}
\end{figure}

In this paper we suggest that the ``other principle'' to obtain the inhomogeneous Maxwell's equations mentioned by Feynman may be the  principle of local charge conservation represented by the continuity equation. We show how a heuristic procedure involving formal operations on the continuity equation evaluated at the retarded time leads to a first-order equation in which we identify the retarded scalar and vector potentials. We then apply the D'Alembertian operator to the retarded potentials, obtaining the wave equations they satisfy. In the final step, we use these wave equations to get not only the inhomogeneous Maxwell's equations but also the homogeneous ones. Our approach is axiomatic in the sense that it starts with the continuity equation as the basic axiom but it is also heuristic in the sense that this equation is heuristically handled. We also show that this axiomatic-heuristic procedure to obtain the full set of Maxwell's equations can be covariantly developed in the Minkowski spacetime. To put in context our axiomatic-heuristic procedure,  it is pertinent to mention that in a series of papers \cite{6,7,8} which originated some comments \cite{9,10} and their respective replies \cite{11,12}, one of us has developed the idea of getting Maxwell's equations by starting with the continuity equation evaluated at the retarded time but without appealing to potentials as we now do in the present paper. In the cited papers it has been argued that charge conservation and causality, respectively represented by the local continuity equation and the retarded time are the cornerstones on which Maxwell's equations are based and therefore they can be considered to be the two fundamental postulates for these equations. It is worth mentioning that although Maxwell's equations are universally accepted, the question of what their fundamental physical postulates are remains a topic of discussion and debate \cite{13,14,15,16,17,18,19}.

The derivation of Maxwell's equations presented here in its three-dimensional and four-dimensional versions, which considers potentials as primary quantities and fields as derived quantities, may be useful to grasp the background of Maxwell's theory and may be presented in undergraduate courses of electromagnetism.

\section*{{{\fontfamily{qag}\selectfont {\large \textcolor[rgb]{0.00,0.00,0.49}{2. Introducing the potentials $\Phi$ and \textbf{A} before the fields \textbf{E} and\textbf{ B}}}}}}
The electric charge conservation can locally be expressed by the continuity equation
\begin{equation}
\nabla\cdot\!\textbf{ J}+\frac{\partial\rho}{\partial t}=0,
\end{equation}
where $\rho$ and $\textbf{J} $ are the\emph{ localised }charge and current densities\footnote[3]{By localised we mean that the densities $\rho$ and $\textbf{J} $ are zero outside a finite region of space. We note that $\rho$ and $\textbf{J}$ could be also non-localised sources. But in this case they should vanish sufficiently rapidly at spatial infinity so that the surface integrals involving these sources vanish at infinity.} which are functions of space and time. Our approach to obtain Maxwell's equations involves two ingredients: The basic axiom expressed by the continuity equation and a heuristic handle of this equation which involves the concept of causality.

We then assume the existence of certain functions of space and time which are \emph{causally} produced by these localised charge and current densities. Let us call these other unknown functions ``the potentials.'' We will justify this name later. We additionally assume that these potentials vanish sufficiently rapidly at spatial infinity so that the surface integrals containing these potentials vanish at infinity. Our first task will consist in finding the explicit form of these unknown potentials and the equations they satisfy. The causal connection between the expected potentials and their sources $\rho$ and $\textbf{J}$ means that the latter precede in time to the former, i.e., the potentials calculated at the field point $ \textbf{x}$ at the time $t$ are caused by the action of their sources $\rho$ and $\textbf{J}$ a distance $R=|\textbf{x}-\textbf{x}'|$ away at the source point $\textbf{x}'$ at the retarded time $t'\!=\!t -t_0$. It is clear that $t_0\!>\!0$ is the time required for the carrier of the charge-potential connection to travel the distance $R$ between the source point $\textbf{ x}'$ and the point $ \textbf{x}$. Consider now that the carrier of the interaction is the photon which moves in a straight line at the speed of light $c$ in vacuum. This implies $t_0=R/c$ and thus the retarded time takes the form:  $t'\!=t -R/c$. Put differently, causality demands that the unknown potentials must be determined by their sources  $\rho$ and $\textbf{J}$ evaluated at the retarded time. We then enclose the terms of the left of (1) in the retardation symbol  $[\;\;]$ which indicates that the enclosed quantity is to be evaluated at the source point
$\textbf{x}'$ at the retarded time $t'=t-R/c$,\footnote[4]{A nice interpretation of equation (2) has been given in Ref.~\cite{6}: ``Consider
an observer at a particular location in space who has a watch
that reads a particular time. The observer is surrounded by
nested spheres, on each of which there is a well-defined retarded
time (with respect to the observer). Equation (2) states
that the continuity equation holds (or rather, held) on each of
those spheres, at the relevant retarded time.''}
\begin{equation}
[\nabla'\cdot\!\textbf{ J}]+\bigg[\frac{\partial\rho}{\partial t'}\bigg]=0.
\end{equation}
We now multiply the first term of (2) by the factor $\mu_0/(4\pi R)$ and the second term of (2) by the equivalent factor
$1/(4\pi\epsilon_0 R c^2)$,\footnote[5]{The presence of the $1/R$ pieces in these factors is consistent with our assumption that the envisioned  potentials vanish sufficiently rapidly at spatial infinity and guarantee that such potentials are uniquely determined.} where $\epsilon_0$ and $\mu_0$ are constants satisfying the relation $\epsilon_0\mu_0\!=\!1/c^2$, and integrate over all space, obtaining the equation
\begin{equation}
\frac{\mu_0}{4\pi}\!\int\!\frac{[\nabla'\cdot\!\textbf{ J}]}{R}d^3x' + \frac{1}{c^2}\frac{1}{4\pi\epsilon_0}\!\int\! \frac{[\partial\rho/\partial t']}{R}d^3x'=0.
\end{equation}
With the idea of taking out the derivative operators from the integrals in  (3), we perform an integration by parts in the first term of (3), in which we use the result \cite{20}: $[\nabla'\cdot\!\textbf{ J}]/R=\nabla\cdot([\textbf{ J}]/R)+ \nabla'\cdot([\textbf{ J}]/R)$ and the fact that the surface integral arising from the term $\nabla'\cdot([\textbf{ J}]/R)$ vanishes at spatial infinity because $\textbf{J}$ is localised. Next we use the result \cite{6}:  $[\partial\rho/\partial t']\!=\!\partial[\rho]/\partial t$ in the second term of (3). After performing the specified operations, the final result reads
\begin{equation}
\nabla\cdot\Bigg\{\frac{\mu_0}{4\pi}\!\int\!\frac{[\textbf{ J}]}{R}d^3x'\Bigg\} + \frac{1}{c^2}\frac{\partial}{\partial t}\Bigg\{\frac{1}{4\pi\epsilon_0} \!\int\!\frac{[\rho]}{R}d^3x'\Bigg\} =0.
\end{equation}
The terms within the curly braces $\{...\}$  are determined by the retarded values of the sources $\textbf{J}$ and $\rho$. We call these terms the retarded vector potential $\textbf{A}$ and the retarded scalar potential $\Phi$:
\begin{equation}
\textbf{A}=\frac{\mu_0}{4\pi}\!\int\!\frac{[\textbf{ J}]}{R}d^3x', \quad \Phi=\frac{1}{4\pi\epsilon_0} \!\int\! \frac{[\rho]}{R}d^3x'.
\end{equation}
These are the potentials we were looking for. Thus, equation (4) takes the compact form
\begin{equation}
\nabla\cdot \textbf{A} + \frac{1}{c^2}\frac{\partial\Phi}{\partial t}=0.
\end{equation}
In the standard presentation of Maxwell's equations, the relation (6) is interpreted as a gauge condition, the so-called Lorenz condition. In our presentation, equation (6) should be rather interpreted as a field equation for potentials. At this stage we wonder what other field equations satisfy the potentials \textbf{A} and $\Phi$.
We then apply the d'Alembert operator $\Box^2\!\equiv\!\nabla^2\!-\!(1/c^2)\partial^2/\partial t^2$ to the potentials in  (5), use the result \cite{6}:
\begin{equation}
\square^2\bigg\{\frac{[{\cal F}]}{R}\bigg\}\!=\!-4\pi[{\cal F}]\delta(\textbf{x}\!-\!\textbf{x}'),
\end{equation}
where ${\cal F}$ is a function of space and time and $\delta$ is the Dirac delta function,\footnote[6]{Equation (7) is proved in reference \cite{6}. As pointed out in reference \cite{18} this identity is true for functions ${\cal F}$ such that the quantities $[{\cal F}]/R$ have not the form  $[{\cal F}]/R=f(R)[\textbf{F}]$. If for example $f(R)=R$ then $[{\cal F}]/R=R[\textbf{F}]$. It follows that $\Box^2 (R[\textbf{F}])=0$ since the quantity $-4\pi R[\textbf{F}]\delta(\textbf{x}-\textbf{x}')$ vanishes for $\textbf{x}\not=\textbf{x}'$ because of the delta function and also for $\textbf{x}=\mathbf{x}'$  because this equality implies $R=0$.} and finally integrate the resulting expressions over all space. After this calculation, we get two wave equations
\begin{equation}
\Box^2 \textbf{A}=-\mu_0\textbf{J},\quad   \Box^2 \Phi=-\frac{\rho}{\epsilon_0}.
\end{equation}
These are the second-order equations we were looking for. They imply expressions for the charge and current densities: $\textbf{J}\!=\!-\,\Box^2 \textbf{A}/\mu_0$ and  $\rho\!=\!-\,\epsilon_0\Box^2\Phi$ that satisfy the continuity equation
\begin{equation}
\nabla\cdot\!\textbf{ J}+\frac{\partial\rho}{\partial t}=-\frac{1}{\mu_0} \Box^2\Bigg(\nabla\cdot \textbf{A} + \frac{1}{c^2}\frac{\partial\Phi}{\partial t}        \Bigg)=0,
\end{equation}
because of (6). The retarded potentials $\textbf{A}$ and $\Phi$ given in (5) constitute the causal solution of the set formed by equations (6) and (8). This solution  is shown to be unique \cite{21}.\footnote[7]{Using relativistic notation, Anderson in Ref.~ \cite{21} has shown that the retarded potentials satisfy a set of sufficient initial and boundary conditions to guarantee their uniqueness. It is clear that our approach to Maxwell's equations does not consider the gauge symmetry because we have constructed retarded potentials with the property to be unique. In other words: we have no gauge freedom in our approach to Maxwell's equations.}

Equations (8) form a set of \emph{second order} equations connecting the potentials $\textbf{A}$ and $\Phi$ with their sources $\textbf{J}$ and $\rho$. A question arises: could there be a set of \emph{first order} field equations equivalent to the set of equations in (8)? Let us investigate this possibility. Using the identity $\nabla^2\textbf{A}\equiv  \nabla (\nabla\cdot \textbf{A}) -\nabla \times (\nabla \times \textbf{A})$ and (6) and (8) we get the equivalent system of equations
\begin{eqnarray}
\nabla\cdot\bigg\{\!-\nabla\Phi-\frac{\partial\textbf{A}}{\partial t}\bigg\} = \frac{\rho}{\epsilon_0}, \\
\nabla\times\big\{\nabla\times\textbf{A}\big\} -\frac{1}{c^2}\frac{\partial}{\partial t}\bigg\{\!-\nabla\Phi-\frac{\partial\textbf{A}}{\partial t}\bigg\} =\mu_0 \textbf{J}.
\end{eqnarray}
 We realise that the quantity $\{-\nabla\Phi-\partial\textbf{A}/\partial t\}$ appears in both (10) and (11), and this does not seem to be a fortuitous coincidence. This quantity together with its partner $\{\nabla\times\textbf{A}\}$ could be physically significant. Let us introduce the fields $\textbf{E}$ and $\textbf{B}$ through the equations
\begin{eqnarray}
\textbf{E}=-\nabla\Phi-\frac{\partial\textbf{A}}{\partial t},\quad \textbf{B}= \nabla\times\textbf{A}.
\end{eqnarray}
This justifies the name of potentials to the functions \textbf{A} and $\Phi$. According to (12) these potentials determine the fields $\textbf{E}$ and $\textbf{B}$.
In terms of \textbf{E} and \textbf{B}, (10) and (11) take the compact form
\begin{eqnarray}
\nabla\cdot \textbf{E} = \frac{\rho}{\epsilon_0}, \\
\nabla\times \textbf{B} -\frac{1}{c^2}\frac{\partial \textbf{E}}{\partial t} =\mu_0 \textbf{J}.
\end{eqnarray}
We note that the divergence of (14) together with  (13) yield  (1) back. Clearly, we have inferred other equivalent expressions for $\textbf{J}$ and $\rho$, namely, $\textbf{J}=\nabla\times \textbf{B}/\mu_0 -\epsilon_0\partial \textbf{E}/\partial t $ and $\rho=\epsilon_0\nabla\cdot \textbf{E}$, which satisfy the continuity equation (1). Of course, (13) and (14) must be completed with other two equations that specify the quantities $\nabla\cdot \textbf{B}$ and $\nabla\times \textbf{E}$ as dictated by the Helmholtz theorem \cite{22}. These other equations are not difficult to find. We quickly note that the fields $\textbf{E}$ and $\textbf{B}$ given in  (12) imply the other two field equations
\begin{eqnarray}
\nabla\cdot \textbf{B}=0, \\
\nabla\times \textbf{E} +\frac{\partial \textbf{B}}{\partial t} =0.
\end{eqnarray}
The set formed by the first order equations (13)-(16) is equivalent to the set formed by the second order equations (8) together with the equation (6). The set of equations (13)-(16) is uniquely determined whenever we adopt boundary conditions for the fields $\textbf{E}$ and $\textbf{B}$ that are consistent with those of the potentials $\textbf{A}$ and $\Phi$.

In order to find the significance of the fields $\textbf{E}$ and $\textbf{B}$ we use (5) and (12) and obtain the retarded solutions of (13)-(16),
\begin{eqnarray}
\textbf{E}=-\nabla \!\int\! \frac{[\rho]}{4\pi\epsilon_0R}d^3x' -\frac{\partial}{\partial t}\!\int\!\!\frac{[\textbf{J}]}{4\pi\epsilon_0 c^2 R}d^3x',\\
\textbf{B}=\nabla\times\!\int\!\!\frac{[\textbf{J}]}{4\pi\epsilon_0 c^2 R}d^3x'.
\end{eqnarray}
It becomes evident that $\textbf{E}$ and $\textbf{B}$ are retarded fields. The system formed by the \emph{coupled} four first-order equations (13)-(16) imply a system formed by two \emph{uncoupled} second-order equations. To find the latter system we apply the d'Alembertian operator $\Box^2$ to (17) and (18), use (7), and integrate the resulting expressions over all space to get the wave equations
\begin{equation}
\Box^2 \textbf{E}=\frac{1}{\epsilon_0}\nabla\rho +\mu_0\frac{\partial\textbf{J}}{\partial t},\quad   \Box^2\textbf{B} =-\mu_0\nabla\times\textbf{J}.
\end{equation}
Our task will be complete if we identify $\epsilon_0$ and $\mu_0$ with the vacuum permittivity and the vacuum permeability. With this identification, the potentials $\Phi$ and \textbf{A} are the electromagnetic scalar and vector potentials and the fields $\textbf{E}$ and $\textbf{B}$ are the electric and magnetic fields.

We have obtained two equivalent versions of electromagnetic field equations. The first one is represented by equations (6) and (8) which are expressed in terms of the retarded scalar and vector potentials defined in (5) and the second one is represented by equations (13)-(16) which are expressed in terms of the retarded fields defined by equations (17) and (18). This second version of the equations is identified with the familiar Maxwell's equations.

Let us emphasize that the fundamental elements of our axiomatic-heuristic approach to find the Maxwell equations were the principle of charge conservation expressed by the continuity equation (the basic axiom) and an heuristic handle of this equation which involved the principle of causality
represented by the retarded time.

\section*{{{\fontfamily{qag}\selectfont {\large \textcolor[rgb]{0.00,0.00,0.49}{3. Introducing the four-potential $\textbf{A}^{\bfmu}$ before the electromagnetic field  $\textbf{F}^{\bfmu\bfnu}$}}}}}

The preceding axiomatic-heuristic approach can also be used to obtain the Maxwell equations in the four-dimensional Minkowski spacetime. Let us introduce the corresponding notation. A point is denoted by $x=x^{\mu}=\{x^0, x^i\}=\{ct, \textbf{x}\}$ and the signature of the metric is $(+,-,-,-).$ Greek indices run from 0 to 3 and  Latin indices run from 1 to 3. The summation convention on repeated indices is adopted.

The continuity equation in the four-space is elegantly simple
\begin{eqnarray}
\partial_\nu J^\nu =0,
\end{eqnarray}
where $J^\nu$ is the four-current which is assumed to be a localised function of spacetime and $\partial_\mu$
is the four gradient. Our basic axiom is now represented by the covariant form of the continuity equation. A heuristic manipulation of this equation will lead us to the manifestly covariant form of Maxwell's equations.

Our first task consists in finding a four-potential which is \emph{causally} connected with the four-current via a covariant equation. The causal connection will be now implemented through the retarded Green function
$G\!=\!G(x,x')$ for the four-dimensional wave equation: $\partial_\mu\partial^\mu G\!=\!\delta^{(4)}(x\!-\!x'),$ where $\partial_\mu\partial^\mu\!=\!-\Box^2$ is the wave operator and $\delta^{(4)}(x\!-\!x')$ is the four-dimensional delta function.
Integration of this wave equation yields the explicit form: $G\!=\!\delta\{t'\!-t+R/c\}/(4\pi R)$.\footnote[8]{A heuristic way to get this Green function is discussed in Ref.~\cite{8}. Notice that this retarded form of the function $G$ is not explicitly Lorentz-invariant. An equivalent form of the function $G$ which is Lorentz-invariant is given by $D_r(x,x')\!=\!\Theta(x_0\!-\!x'_0)\delta[(x\!-\!x')^2]/(2\pi),$ where $\Theta$ is the theta function. See Ref.~\cite{23}.} The function $G$ satisfies the property $\partial^\mu G\!=\!-\partial'^\mu G.$ We now evaluate (20) at the source point $x'$ and multiply the resulting equation by $\mu_0G$ and integrate over all spacetime, obtaining
\begin{eqnarray}
\int\! \mu_0G\partial'_\nu J^\nu d^4x'=0.
\end{eqnarray}
After an integration by parts in  (21), in which we use the relation $G\partial'_\nu J^\nu\!=\! \partial_\nu (G J^\nu) + \partial'_\nu (G J^\nu)$ and the fact that the surface integral originated by the term $\partial'_\nu (G J^\nu)$ vanishes at spatial infinity, we take out the operator $\partial_\nu$ from the integral in (21) and obtain
\begin{eqnarray}
\partial_\nu\!\! \int\!\mu_0 G J^\nu d^4x'=0.
\end{eqnarray}
The integral in (22) must have some significant interpretation, we call it the four-potential
\begin{eqnarray}
A^\nu\!= \!\mu_0\!\int\!\!G J^\nu d^4x',
\end{eqnarray}
in terms of which (22) becomes elegantly simple compact
\begin{eqnarray}
\partial_\nu A^\nu=0.
\end{eqnarray}
In the next step we take the wave operator $\partial_\mu\partial^\mu$ to  (23), use  the result $\partial_\mu\partial^\mu G\!=\!\delta^{(4)}(x-x')$ and integrate over all spacetime to obtain the wave equation
\begin{eqnarray}
\partial_\mu\partial^\mu A^\nu=\mu_0J^\nu.
\end{eqnarray}
This is the covariant equation we were looking for. It clearly provides an expression for the four-current $J^\nu=\partial_\mu\partial^\mu A^\nu/\mu_0$ that satisfies the continuity equation
\begin{eqnarray}
\partial_\nu J^\nu=\frac{1}{\mu_0}\partial_\mu\partial^\mu \partial_\nu A^\nu=0,
\end{eqnarray}
because of (24). Equation (25) is a second-order equation that causally connects the four-potential  $A^\nu$ with the four-current   $J^\nu$.
Are there two first-order equations equivalent to the equation (25)? The answer is in the affirmative. We combine (24) and (25) to get the equation
\begin{eqnarray}
\partial_\mu\big\{\partial^\mu A^\nu-\partial^\nu A^\mu\big\}=\mu_0J^\nu.
\end{eqnarray}
We strongly suspect that the antisymmetric tensor $\partial^\mu A^\nu-\partial^\nu A^\mu$ could be physically significant. We  find convenient to label this antisymmetric tensor as
\begin{eqnarray}
F^{\mu\nu}=\partial^\mu A^\nu-\partial^\nu A^\mu,
\end{eqnarray}
in terms of which (27) takes the elegant form
\begin{eqnarray}
\partial_\mu F^{\mu\nu} =\mu_0J^\nu.
\end{eqnarray}
This provides us another expression for the four-current $J^\nu=\partial_\mu F^{\mu\nu}/\mu_0$ that satisfies the continuity equation
\begin{eqnarray}
\partial_\nu J^\nu=\frac{1}{\mu_0}\partial_\mu\partial_\nu  F^{\mu\nu}=0,
\end{eqnarray}
because $\partial_\mu\partial_\nu  F^{\mu\nu}\equiv 0$ since the operator $\partial_\mu\partial_\nu$ is symmetric in the indices $\mu$ and $\nu$ and the tensor $F^{\mu\nu}$ is antisymmetric in these indices. On the other hand, any antisymmetric tensor field $F^{\mu\nu}$ in the four-space has an associated a dual tensor defined by
$^*\!{F}^{\mu\nu}\!=\!(1/2)\varepsilon^{\mu\nu\alpha\beta}F_{\alpha\beta}$, where $\varepsilon^{\mu\nu\alpha\beta}$ is the four-dimensional Levi-Civita symbol with $\varepsilon^{0123}\!=\!1$. A generalised Helmholtz theorem \cite{22,24} states that an antisymmetric tensor field is completely determined by specifying its divergence and the divergence of its dual. We can show that the dual of (28) is given by  $ ^*\!{F}^{\mu\nu}\!=\!\varepsilon^{\mu\nu\alpha\beta}\partial_{\alpha}A_\beta$ and its divergence reads $\partial_\mu\,\!^*\!{F}^{\mu\nu}\!=\!\varepsilon^{\mu\nu\alpha\beta}\partial_\mu\partial_{\alpha}A_\beta,$ whose right-hand side identically vanishes
because $\varepsilon^{\mu\nu\alpha\beta}$ is  antisymmetric in the indices $\mu$ and $\alpha$ and the operator $\partial_\mu\partial_\alpha$ is symmetric in these indices. Therefore the additional required field equation is given by
\begin{eqnarray}
\partial_\mu\,\!^*\!{F}^{\mu\nu}=0.
\end{eqnarray}
The set formed by equations (24) and (25) is equivalent to the set formed by equations (29) and (31). Let us write (28) as
$F^{\mu\nu} =(\delta^\nu_\lambda\partial^\mu-\delta^\mu_\lambda\partial^\nu) A^\lambda$, were $\delta^\nu_\lambda$ is the Kronecker delta. Using this expression for $F^{\mu\nu}$ together with (23) we obtain
\begin{eqnarray}
F^{\mu\nu}=\mu_0 (\delta^\nu_\lambda\partial^\mu-\delta^\mu_\lambda\partial^\nu) \!\int\!\!G J^\lambda d^4x'.
\end{eqnarray}
We now take the wave operator  $\partial_\alpha\partial^\alpha$ to  (32), use $\partial_\mu\partial^\mu G\!=\!\delta^{(4)}(x-x')$ and integrate over all spacetime, obtaining the wave equation
\begin{eqnarray}
\partial_\alpha\partial^\alpha F^{\mu\nu}=\mu_0 (\partial^\mu J^\nu-\partial^\nu J^\mu).
\end{eqnarray}
Our task will be complete if we appropriately specify the components of the four-current $J^{\mu}$, the four-potential $A^{\mu}$, the electromagnetic field $F^{\mu\nu}$ and its dual $^*\!{F}^{\mu\nu}$. The four-gradient is defined by $\partial_\mu=\big\{(1/c)\partial/\partial t, \nabla\big\}$. Therefore, if we write
\begin{eqnarray}
  J^\nu\!=\!\{c\rho, \textbf{J}\}, \quad A^\nu\!=\!\{\Phi/c, \textbf{A}\},
\end{eqnarray}
then (23)-(25) reproduce (5), (6) and (8) respectively. Similarly, if we write
\begin{eqnarray}
F^{i0}\!=\!(\textbf{E})^i/c,\quad F^{ij}\!=\!-\varepsilon^{ijk}(\textbf{B})_k,\quad ^*\!{F}^{i 0}\!=\!(\textbf{B})^i,\quad ^*\!{F}^{i j}\!=\!\varepsilon^{ijk}( \textbf{E})_k/c.
\end{eqnarray}
 where $(\textbf{E})^i$ and $(\textbf{B})_k$ are the Cartesian components of the fields \textbf{E} and \textbf{B}, then (29), (31) and (32) reproduce (13)-(18).

We have obtained two equivalent covariant versions of the electromagnetic field equations in the Minkowski spacetime. The first one is represented by equations (24) and (25) which are expressed in terms of the retarded four-potential defined in (23). The second one is represented by equations (29) and (31) which are expressed in terms of the retarded electromagnetic field (32) and its dual. This second version of the equations is identified with the covariant form of Maxwell's equations. The basic physical ingredients of our axiomatic-heuristic procedure to find these equations were charge conservation mathematically represented by the covariant form of the continuity equation and a heuristic handling of this equation involving the retarded Green function of the wave equation.

\section*{{{\fontfamily{qag}\selectfont {\large \textcolor[rgb]{0.00,0.00,0.49}{4. Discussion}}}}}

How should we interpret the procedure proposed here to \emph{obtain} Maxwell's equations?
Have we really made a \emph{derivation} of these equations or just a \emph{construction} of them?

Following the traditional procedure starting with Maxwell's equations, one introduces potentials and derives their wave equations (by adopting the Lorenz condition). By assuming appropriate boundary conditions the solutions of these wave equations yield the retarded potentials which are then differentiated to get the corresponding retarded electric and magnetic fields. This conventional procedure is logically well-structured and then one can conclude that if Maxwell's equations are postulated from the beginning then one can derive the retarded potentials and hence their corresponding fields.  End of the story.

On the other hand, the reverse procedure starting with the retarded potentials and ending with Maxwell's equations does not seem to be simple at first sight. Suppose that by some means (which  of course does not involve the Maxwell equations) we have found the retarded potentials (5). Differentiating these potentials one obtains their wave equations (8) and equation (6). Combining (6) and (8) one infers equations (10) and (11) which are then identified with the inhomogeneous Maxwell's equations whenever the electric and magnetic fields are defined as (12). In the final step, one uses these definitions of fields to obtain the homogeneous Maxwell's equations. This reversed procedure is conceptual and pedagogically significative
as long as one can convincingly justify the existence of the retarded potentials without explicitly appealing to Maxwell's equations. This is the most difficult problem to solve.

But there is a conceptual disadvantage in the traditional procedure. If one \emph{postulates} Maxwell's equations from the beginning then the task of identifying the basic postulates of these equations \emph{loses} its meaning. On the contrary, the reversed procedure
starting with retarded potentials can help to elucidate the nature of these postulates. In the task of finding these potentials, we have
argued that charge conservation should be considered the fundamental axiom underlying Maxwell's equations.

Clearly, the interest sketched by Feynman in his handwritten notes was how to obtain Maxwell's equations by starting with potentials and using physical principles like relativity and charge conservation. In this aim we think the recourse of heuristic arguments is unavoidable. Put differently, the procedure followed by De Luca et al. \cite{3} to arrive at the Lorenz force and the homogeneous Maxwell's equation as well as our procedure to arrive at the inhomogeneous and homogeneous Maxwell's equations could be interpreted as \emph{constructive} procedures. \emph{In this kind of procedures one makes use of heuristic arguments to show the \emph{existence} of a mathematical object by providing a method for creating the object.}
 Of course, one generally has knowledge of this object by other means. In this perspective, our procedure to obtain Maxwell's equations could be considered as a constructive method to demonstrate the existence of retarded potentials which leads to the electric and magnetic fields satisfying Maxwell's equations. In other words, from a conceptual point of view our procedure could (and should!) be formulated as an existence theorem. Let us enunciate this theorem.
\vskip 5pt
\noindent \emph{Existence Theorem}. Let $\cJ(\textbf{x},t)$ and $\mathscr{G}(\textbf{x},t)$ be vector and scalar functions which are spatially localised and satisfy the continuity equation
\begin{equation}
\nabla\cdot \cJ+\frac{\partial\mathscr{G}}{\partial t}=0.
\end{equation}
If this equation is evaluated at the source point
$\textbf{x}'$ at the retarded time $t'=t-R/\mathscr{C}$ with $\mathscr{C}$ being a constant with units of velocity,
then \emph{there exist} the retarded scalar and vector functions: ${\mA}(\textbf{x},t)$ and $\mathscr{P}(\textbf{x},t)$ defined by
\begin{equation}
\mA=\frac{1}{4\pi}\!\int\!\frac{[\cJ]}{R}d^3x', \quad \mathscr{P}=\frac{1}{4\pi}\!\int\! \frac{[\mathscr{G}]}{R}d^3x',
\end{equation}
that satisfy the equation
\begin{equation}
\nabla\cdot \mA +\frac{\partial \mathscr{P}}{\partial t}=0,
\end{equation}
where the retardation symbol  $[\;\;]$ indicate that the enclosed quantity is to be evaluated at the source point at the retarded time.

\vskip 5pt
\noindent
\emph{Corollary 1.} The functions $\mathscr{P}$ and ${\mA}$ in (37) satisfy the wave equations
\begin{equation}
\Box^2\mathscr{P} =-\mathscr{G}, \quad \Box^2 \mA=-\cJ,
\end{equation}
where $\Box^2\!\equiv\!\nabla^2\!-\!(1/\mathscr{C}^2)\partial^2/\partial t^2$. 
\vskip 5pt
\noindent
\emph{Corollary 2.} There exist retarded fields: ${\cE}(\textbf{x},t)$ and ${\cB}(\textbf{x},t)$ defined by
\begin{eqnarray}
{\cE}=-\nabla\mathscr{P}-\frac{1}{\mathscr{C}^2}\frac{\partial \mA}{\partial t},\quad {\cB}= \nabla\times \mA,
\end{eqnarray}
that satisfy the field equations
\begin{eqnarray}
\nabla\cdot {\cE} =\mathscr{G} , \quad \nabla\times {\cE} +\frac{1}{\mathscr{C}^2}\frac{\partial {\cB}}{\partial t} =0,\\
\nabla\cdot {\cB}=0, \quad \nabla\times {\cB} -\frac{\partial {\cal E}}{\partial t} =\cJ.
\end{eqnarray}
\noindent The proof of this general theorem and the proof of its corollaries are entirely similar to those given in the section 2 for the particular case of electromagnetic expressions in SI units.\footnote[9]{The formulated theorem is of general character and can be applied to
scalar and vector source functions of theories different from that of Maxwell. However, if we make the specifications
$\mathscr{C}=c,\;\cJ=\textbf{J},\; \mathscr{G}=\rho,\; \mA=\textbf{A}/\beta, \; \mathscr{P}=\Phi/\alpha,\; {\cB}=\textbf{B}/\beta,\;
{\cE}=\textbf{E}/\alpha$ where $\alpha=\beta\chi c^2$ then the theorem describes the Maxwell equations in a form independent of specific units. More precisely, this specification describes Maxwell's equations in the $``\alpha\beta\chi''$ system which involves the Gaussian, SI, and Heaviside-Lorentz unit systems. See Refs.~\cite{6,25}.} Furthermore, if we make the particular specifications
\begin{eqnarray}
\mathscr{C}=c,\;\cJ=\textbf{J},\; \mathscr{G}=\rho,\; \mA=\textbf{A}/\mu_0, \; \mathscr{P}=\epsilon_0\Phi,\; {\cB}=\textbf{B}/\mu_0,\;
{\cE}=\epsilon_0\textbf{E},
\end{eqnarray}
in the general theorem and its corollaries then we obtain the corresponding electromagnetic expressions in SI units. In the Minkowski spacetime the existence theorem is indeed elegant:
\vskip 5pt
\noindent \emph{Existence Theorem}. Let ${\cal J}^\nu$ a localised four-vector that satisfies the continuity equation  $\partial_\nu {\cal J}^\nu =0$ then there exists a four-vector ${\cal A}^\nu$ defined as
\begin{eqnarray}
{\cal A}^\nu =\int {\cal G} {\cal J}^\nu d^4x',
\end{eqnarray}
that satisfies the field equation $\partial_\nu {\cal A}^\nu =0$, where the Green function is defined by ${\cal G}=\delta\{t'\!-t+R/\mathscr{C}\}/(4\pi R)$
with $\mathscr{C}$ being a constant with units of velocity.
\vskip 5pt
\noindent
\emph{Corollary 1.} The four-vector ${\cal A}^\nu$ satisfies the wave equation $\partial_\mu\partial^\mu {\cal A}^\nu={\cal J}^\nu$, where
$\partial_\mu\partial^\mu\!=-\nabla^2\!+\!(1/\mathscr{C}^2)\partial^2/\partial t^2$.
\vskip 5pt
\noindent
\emph{Corollary 2.} There exists the antisymmetric tensor ${\cal F}^{\mu\nu}= \partial^\mu {\cal A}^\nu-\partial^\nu {\cal A}^\mu$ that satisfies
the field equations $\partial_\mu {\cal F}^{\mu\nu} ={\cal J}^\nu$ and $\partial_\mu\,\!^*{\cal F}^{\mu\nu}=0$, where $^*\!{\cal F}^{\mu\nu}\!=\!\varepsilon^{\mu\nu\alpha\beta}\partial_{\alpha}{\cal A}_\beta$.
\vskip 5pt
\noindent The proof of this covariant form of the theorem and the proof of its corollaries are entirely similar to those given in the section 3 for the case of electromagnetic expressions in SI units. If $\mathscr{C}=c$ then  ${\cal G}=G$. If in this case we make ${\cal J}^\nu= J^\nu$ and
${\cal A}^\nu= A^\nu/\mu_0$ with  $A^\nu=(\Phi/c,\mathbf{A})$ then (44) becomes (23) and $A^\nu$ is the electromagnetic four-potential in SI units.

It is possible consider a different heuristic handle of the continuity equation (the basic axiom) to formulate a theorem that is equivalent to
the previously considered existence theorem. For example, we can formulate the following existence theorem \cite{6}: Given the localised sources $\rho(\textbf{x},t)$ and $\textbf{J}(\v x,t)$ satisfying the continuity equation $\nabla\cdot \textbf{J}+\partial\rho/\partial t=0$ there exist the retarded fields $\textbf{F}(\textbf{x},t)$ and $ \textbf{G}(\textbf{x},t)$ defined by
\begin{eqnarray}
 \textbf{F} =
 \frac{\alpha}{4\pi}\int\bigg(\frac{\hat{\textbf{R}}}{R^2}[\rho]+\frac{\hat{\textbf{R}}}{Rc}\left[\frac{\partial \rho}{\partial t}\right]
-\frac{1}{Rc^2}\left[\frac{\partial \textbf{J}}{\partial t}\right]\bigg)\, d^3x',\\
 \textbf{G}=
 \frac{\beta}{4\pi}\int \bigg([\textbf{J}]\times\frac{\hat{\textbf{R}}}{R^2 }+\bigg[\frac{\partial \v J}{\partial t}\bigg]\times\frac{\hat{\textbf{R}}}{R c}\bigg)\, d^3x'.
\end{eqnarray}
that satisfy the following field equations: $\nabla\cdot \textbf{F}=\alpha\rho,\,\nabla\cdot \textbf{G}=0,\,
 \nabla\times\textbf{F}+\chi \partial \textbf{G}/\partial t=0$ and $
\nabla\times \textbf{G}-(\beta/\alpha)\partial \textbf{F}/\partial t
=\beta \textbf{J}.$ Here $\hat{\textbf{R}}={\textbf{R}}/R =(\textbf{x}-\textbf{x}')/|\textbf{x}- \textbf{x}'|$ and equations (45) and (46) are in the $``\alpha\beta\chi$'' system defined by $\alpha=\beta\chi c^2$. In this case the axiomatic-heuristic approach shows the existence of the electric and magnetic fields in the generalized form of Coulomb and Biot-Savart laws given by Jefimenko \cite{6} which satisfy Maxwell's equations.

Similarly, an alternate heuristic manipulation of the continuity equation in the Minkowski spacetime leads to the existence of an electromagnetic tensor satisfying the covariant form of Maxwell's equations. This is a consequence the following existence theorem \cite{7}:   Given the localized four-vector ${\cal J}^{\mu}$ satisfying the continuity equation $\partial_\mu {\cal J}^\mu=0$ there exists the antisymmetric tensor field
\begin{equation}
{\cal F}^{\mu\nu}=
\int  {\cal G}(\partial'^\mu {\cal J}^\nu-\partial'^\nu {\cal J}^\mu)\,d^4x',
\end{equation}
that satisfies the field equations: $\partial_{\mu} {\cal F}^{\mu\nu} = {\cal J}^\nu$ and $
\partial_{\mu}{\cal^*F}^{\mu\nu} = 0,$ where $^*{\cal F}^{\mu\nu}=(1/2)\varepsilon^{\mu\nu\alpha\beta} {\cal F}_{\alpha\beta}$ is the dual of ${\cal F}^{\mu\nu}$ and ${\cal G}=\delta\{t'\!-t+R/\mathscr{C}\}/(4\pi R)$
with $\mathscr{C}$ being a constant whose units are of velocity. If we make the identification $\mathscr{C}=c$ then  ${\cal G}=G$ and if in addition we make ${\cal J}^\mu=\mu_0J^\nu$ with $J^\nu=(c\rho,\textbf{J})$ then ${\cal F}^{\mu\nu}={F}^{\mu\nu}$ is the electromagnetic field tensor in SI units.

The point to remark is that in the proof of an existence theorem of an object, one is generally free to use all heuristic devices that allows one to exhibit the explicit form of such an object. This is the more essential aspect in a constructive approach.

\section*{{{\fontfamily{qag}\selectfont {\large \textcolor[rgb]{0.00,0.00,0.49}{5. On the postulates of Maxwell's equations}}}}}

 Most authors agree that the continuity equation is a \emph{consequence} of Maxwell's equations \cite{26}. Other authors state that it is an \emph{integrability condition} of these equations \cite{27,28}. Some other authors are more cautious and claim that Maxwell's equations are \emph{consistent} with the continuity equation \cite{23,29}. Although Maxwell's equations formally imply the continuity equation, the idea that the latter is a consequence of the former is in a sense questionable. The fact is that the continuity equation has its own existence \emph{independent} of Maxwell's equations. This can be illustrated by the fact that there are field equations of different electromagnetic theories that are also consistent with the continuity equation. For example, one of these theories arises when the Faraday induction term of Maxwell's equations is eliminated, obtaining the field equations of a Galilean-invariant instantaneous electrodynamics \cite{30,31}. Other examples are the Proca equations of the massive electrodynamics \cite{32} and the field equations of an electrodynamics in an Euclidean four-space \cite{33,34}. Therefore, one should interpret the continuity equation as a formal representation of the principle of charge conservation, but having always in mind that this principle is not exclusive of Maxwell's theory.

 Accordingly, we can equally use the continuity equation to formulate other existence theorems for potentials or fields which can be applied to the aforementioned alternative electromagnetic theories. Here we have evaluated this equation at the retarded time to obtain Maxwell's equations. But we can equally evaluate this equation at present time, for example, and following a similar heuristic procedure we will obtain the field equations of a Galilean-invariant instantaneous electrodynamics in Gaussian units \cite{30,31}: $\nabla\cdot \textbf{E}=4\pi \rho,\nabla\cdot \textbf{B}=\!0, \nabla\times\textbf{E}=0$ and $
\nabla\times \textbf{B}-(1/c)\partial \textbf{E}/\partial t
=(4\pi/c) \textbf{J}.$ However, this does not prevent us to consider that the continuity equation is the cornerstone on which Maxwell's equations can be constructed. It is in this sense that we claim that charge conservation must be unavoidable considered as one of the basic postulates of Maxwell's equations. It has been argued that the other basic postulate may be the principle of causality
\cite{6,7,8} represented by the retarded time or by the retarded Green function of the wave equation. Of course, we can integrate these two postulates in a single fundamental postulate which would state that \emph{the continuity equation is valid at all times}. Therefore, evaluating this equation at a particular time is not a new postulate but only one special case of the fundamental postulate.

The alert reader might argue that if charge conservation is really
the fundamental physical principle underlying Maxwell's equations then one should be able to obtain these equations using only the continuity equation without making any further assumptions. In our opinion this demand is very hard to satisfy, at least at the level in which we call basic postulates in physics. Furthermore, as already pointed out, the continuity equation may imply other fields equations depending on the ``further assumptions.''
Let us give an example to illustrate our point. Most physicists would agree that the basic postulates used to derive the Lorentz transformations are
the principle of relativity (the first postulate), which states that physical laws must exhibit the same form in inertial frames, and the constancy of the speed of light (the second postulate), which states that the speed of light is the same in inertial frames. What is not well-known is that in 1887,  Voigt \cite{35} used these same two postulates and derived a set of spacetime transformations different from the Lorentz transformations \cite{36}. In other words, the same postulates may lead to distinct space-time theories! The explanation is simple, the basic postulates are the same but there are different additional assumptions (implicit or explicit) underlying in the derivation of Lorentz and Voigt transformations. We think such additional assumptions are important but they do not qualify to be fundamental postulates.

Similarly, charge conservation can be seen as a basic postulate which requires of some additional considerations to imply Maxwell's equations. One of these additional assumptions is, for example, the retarded time or the retarded Green function of the wave equation. Nevertheless, we should point out that this assumption is sufficient but not necessary since we could equally assume the advanced time $(t'=t+R/c)$ or the advanced Green function of the wave equation $(G\!=\!\delta\{t'\!-t-R/c\}/(4\pi R))$ and obtain Maxwell's equations as well. Put differently, charge conservation is a basic postulate (fully justified by experimental considerations) and causality (represented by the retarded time or the retarded Green function of the wave equation) is a sufficient but not a necessary assumption which --we think-- does not qualify to be a basic postulate but rather as a complementary assumption. Under this wisdom, the idea of considering that charge conservation is the basic postulate of Maxwell's equations is similar to the idea of considering that the  principle of relativity and the constancy of the speed of light are the basic postulates of special relativity.
\section*{{{\fontfamily{qag}\selectfont {\large \textcolor[rgb]{0.00,0.00,0.49}{6. Concluding remarks}}}}}

We have evidence that Feynman attempted to find a different derivation of Maxwell's equations in at least two periods of his life. The first attempt was around 1948, year in which Feynman showed Dyson an unusual proof of the homogeneous Maxwell's equations \cite{37}. Dyson reconstructed Feynman's proof as an existence theorem: If a non-relativistic particle satisfies Newton's law of motion and the commutation relations between its position and velocity then \emph{there exist} two fields that satisfy the Lorentz force and the homogeneous Maxwell's equations. The inhomogeneous Maxwell's equations were merely assumed to be the definitions of charge and current densities. The second attempt was at the end of 1963 as may be seen in the Feynman's handwritten notes recently discovered by Gottlieb \cite{2} and discussed by De Luca et al \cite{3}. In this second attempt, the Lorentz force was inferred by assuming that the force that acts on a charge is linear in its velocity and is the spatial component of a four-force of special relativity. The homogeneous Maxwell's equations were obtained via the well-known principle of least action. There is a certain parallelism between these two attempts: both were unpublished and both fail to obtain the inhomogeneous Maxwell's equations. In the first attempt these equations were defined but not derived and in the second attempt they were not inferred. Charge conservation represented by the continuity equation was not considered in both attempts. Perhaps  we may never know what Feynman had in mind in 1966 when he said that he had ``cooked up a much better way of presenting the electrodynamics, a much more original and much more powerful way than is in the book,'' but it is intriguing that in his first handwritten page wrote in 1963 (see figure 1) he clearly wrote charge conservation and not charge invariance. Was this an error or an unconscious desire?

Here we have pointed out that charge conservation expressed by the continuity equation is the key to obtain the Maxwell equations. We have shown that if the continuity equation evaluated at the retarded time is heuristically handled then we can show that there exist defined retarded potentials that imply not only the inhomogeneous  Maxwell's equations but also the homogeneous ones. In the search for this alternative presentation of Maxwell's equations in which potentials are introduced before fields, we have been motivated by Feynman's words that \cite{38}: ``... there is a pleasure in recognising old things from a new point of view. Also, there are problems for which the new point of view offers a distinct advantage.''

\vskip 25pt

\section*{{{\fontfamily{qag}\selectfont {\large \textcolor[rgb]{0.00,0.00,0.49}{Acknowledgment}}}}}

\noindent We dedicate this paper to the memory of Richard P. Feynman\dag \,on the occasion of its 101st anniversary.

\section*{{{\fontfamily{qag}\selectfont {\large \textcolor[rgb]{0.00,0.00,0.49}{References}}}}}

{}


\begin{thebibliography}{37}
\bibitem{1}
R. P. Feynman, R. B. Leighton and M. Sands. The Feynman Lectures on Physics. Addison-Wesley
(1963).


\bibitem{2}
Available from the online \emph{Feynman's Lectures on Physiscs} (\href{http://www.feynmanlectures.caltech.edu/}{www.feynmanlectures.caltech.edu}). See: \href{http://www.feynmanlectures.caltech.edu/info/other/Alternate_Way_to_Handle_Electrodynamics.html}{http://www.feynmanlectures.caltech.edu/info/other/Alternate\_Way\_to\_Handle\_Electrodynamics.html}

\bibitem{3}
De Luca R, Di Mauro M, Esposito S, and Naddeo A 2019 Feynman's different approach to electromagnetism \emph{Eur. J. Phys.} \href{https://doi.org/10.1088/1361-6404/ab423a}{\textbf{40}, 065205}




\bibitem{4}
Feynman R P Lectures on Electrostatics, Electrodynamics, Matter-Waves Interacting, Relativity. Lectures at the Hughes Aircraft Company; notes taken and transcribed by John T. Neer (1967-8)\\
\href{http://www.thehugheslectures.info/wp-content/uploads/lectures/FeynmanHughesLectures_Vol2.pdf}
{http://www.thehugheslectures.info/wp-content/uploads/lectures/FeynmanHughesLecturesVol2.pdf}



\bibitem{5}
Goodstein D and Goodstein J 2000 Richard Feynman and the History of Superconductivity \emph{Phys. Perspect.} \href{https://doi.org/10.1007/s000160050035}{\textbf{2} 3-47}

\bibitem{6}
Heras J A 2007 Can Maxwell's equations be obtained from the continuity equation? \emph{Am. J. Phys.} \href{http://dx.doi.org/10.1119/1.2739570}{\textbf{75} 652-56}


\bibitem{7}
Heras J A 2009 How to obtain the covariant form of Maxwell's equations from the continuity equation \emph{Eur. J. Phys.} \href{http://dx.doi.org/10.1088/0143-0807/30/4/017}{ \textbf{30}, 845-54}


\bibitem{8}
Heras J A 2016 An axiomatic approach to Maxwell's equations \emph{Eur. J. Phys.} \href{http://dx.doi.org/10.1088/0143-0807/37/5/055204}{\textbf{37}, 055204}

\bibitem{9}
Jefimenko O D 2008 Causal equations for electric and magnetic fields and Maxwell's equations: Comment on a paper by Heras \emph{Am. J. Phys.}
 \href{http://dx.doi.org/10.1119/1.2825390}{\textbf{76}, 101}

\bibitem{10}
Kapuscik E 2009 Comment on `Can Maxwell's equations be obtained from the continuity equation?' by Heras J A [Am. J. Phys. \textbf{75} 652-56 (2007)] \emph{Am. J. Phys.} \href{http://dx.doi.org/10.1119/1.3039030}{ \textbf{77}, 754}

\bibitem{11}
Heras J A 2008 Author's response \emph{Am. J. Phys.} \href{http://dx.doi.org/10.1119/1.2826656}{ \textbf{76}, 101-2}


\bibitem{12}
Heras J A 2009 Reply to ``Comment on `Can Maxwell's equations be obtained from the continuity equation?'" by E Kapuscik [Am. J. Phys. \textbf{77}, 754 (2009)] \emph{Am. J. Phys.} \href{http://dx.doi.org/10.1119/1.3039031}{\textbf{77}, 755-56}


\bibitem{13}
Hehl F W and Obukhov Y N 2003 \emph{Foundations of Classical Electrodynamics: Charge, Flux, and Metric} (MA: Birkh\"{a}user)



\bibitem{14}
Diener G, Weissbarth J, Grossmann F and Schmidt R Obtaining Maxwell's equations heuristically \emph{Am. J. Phys.} \href{http://dx.doi.org/10.1119/1.4768196}{\textbf{81} 120$-$23}.


\bibitem{15}
Kosyakov B P 2014 The pedagogical value of the four-dimensional picture: II. Another way of looking at the electromagnetic field
\emph{Eur. J. Phys.} \href{http://dx.doi.org/10.1088/0143-0807/35/2/025013}{\textbf{35}  025013}


\bibitem{16}
Sobouti Y 2015 Lorentz covariance almost implies electromagnetism and more \emph{Eur. J. Phys.} \href{http://dx.doi.org/10.1088/0143-0807/36/6/065036}{ \textbf{36}, 065036}

\bibitem{17}
Hanno E and Nordmark A B 2016 Relativistic version of the Feynman$-$Dyson$-$Hughes derivation of the Lorentz force law and Maxwell's homogeneous equations \emph{Eur. J. Phys.} \href{http://dx.doi.org/10.1088/0143-0807/37/5/055201}{\textbf{37}, 05520}

\bibitem{18}
Heras R 2017 Alternative routes to the retarded potentials \emph{Eur. J. Phys.} \href{https://doi.org/10.1088/1361-6404/aa7f18}{ \textbf{38} 055203}

\bibitem{19}
Kosyakov B 2007 \emph{Introduction to the Classical Theory of Particles and Fields} (Springer-Verlag Berlin Heidelberg)
\bibitem{20}
Jefimenko O D 1989 \emph{Electricity and Magnetism} 2nd edn (Star City, WV: Electrect Scientific)
\bibitem{21}
Anderson J L 1967 \emph{Principles of Relativity Physics} (New York: Academic Press)

\bibitem{22}
Heras R 2016 The Helmholtz theorem and retarded fields, \emph{Eur. J. Phys.} \href{http://dx.doi.org/10.1088/0143-0807/37/6/065204}{\textbf{37} 065204}


\bibitem{23}
Jackson J D 1999 \emph{Classical Electrodynamics} 3rd edn (New York: Wiley)


\bibitem{24}
Heras J A 1990 A short proof of the generalized Helmholtz theorem \emph{Am. J. Phys.} \href{http://dx.doi.org/10.1119/1.16225}{ \textbf{58},  154-55 }.

\bibitem{25} Heras J A and B\'aez G 2009 The covariant formulation of Maxwell's equations expressed in a form independent of specific units {\it Eur. J. Phys.} \href{http://dx.doi.org/10.1088/0143-0807/30/1/003}{{\bf 30}  23-33}

 \bibitem{26}
Griffiths D J 1999 \emph{Introduction to Electrodynamics} 3rd edn (Upper Saddle River, NJ: Prentice-Hall).

\bibitem{27}
Burke W L 1985 \emph{Applied Differential Geometry} (Cambridge: Cambridge University Press).

\bibitem{28}
Betounes D 1998 \emph{Partial Differential Equations for Computational Science} (Springer-Verlag New York).


\bibitem{29}
Zangwill A 2012 \emph{Modern Electrodynamics} (Cambridge: Cambridge University Press).

\bibitem{30}
Jammer M and Stachel J 1980 If Maxwell had worked between Amp\'ere and Faraday: An historical fable with a pedagogical moral \emph{Am. J. Phys.} \href{https://doi.org/10.1119/1.12239}{\textbf{48} 5-7}


\bibitem{31}
Heras J A 2005 Instantaneous fields in classical electrodynamics \emph{Eur. Phys. Lett.} \href{http://dx.doi.org/10.1209/epl/i2004-10318-y}{\textbf{69} 1-7}

\bibitem{32}
Goldhaber A S and Nieto M M 2010 Photon and Graviton Mass Limits \href{https://doi.org/10.1103/RevModPhys.82.939}{Rev. Mod. Phys. \textbf{82} 939-979}

\bibitem{33}
Heras J A 1994 Euclidean electromagnetism in four space: A discussion between God and the Devil \emph{Am. J. Phys.} \href{https://doi.org/10.1119/1.17681}{\textbf{62} 914–16}

\bibitem{34}
Heras J A 2006 The Kirchhoff gauge \emph{Ann. Phys.} \href{https://doi.org/10.1016/j.aop.2005.12.001}{\textbf{321} 1265–73}

\bibitem{35}
Voigt W 1887 \"Uber das Doppler'sche \emph{Princip Nachr. Ges. Wiss.} G\"ottingen \textbf{8} 41-51

\bibitem{36}
Heras R A review of Voigt's transformations in the framework of special relativity arXiv:\href{https://arxiv.org/abs/1411.2559}{1411.2559}

\bibitem{37}
 Dyson F J 1990 Feynman's proof of the Maxwell equations {\it Am. J. Phys.} \href{http://dx.doi.org/10.1119/1.16188}{\textcolor[rgb]{0.00,0.00,1.00}{ {\bf 58}  209-11 }}

\bibitem{38}
Feynman R P 1948 Space-time approach to non-relativistic quantum mechanics \emph{Rev. Mod. Phys.}
\href{https://doi.org/10.1103/RevModPhys.20.367}{\textbf{20} 367–87}




 \end{thebibliography}
 \end{document}